\documentclass[12pt]{JHEP3}
\usepackage{pstricks} 
\usepackage{latexsym,amssymb}
\usepackage{psfrag}
\usepackage{graphicx}
\newcommand{\diag}{\mathrm{Diag}}
\newcommand{\Tr}{\mathrm{Tr}\, }
\newcommand{\Det}{\mathrm{Det}\, }
\newcommand{\im}{\mathrm{Im}}

\newrgbcolor{darkred}{0.7 0 0}
\newrgbcolor{darkgreen}{0 0.7 0 }
\newrgbcolor{darkblue}{0 0 0.7}

\title{Neutrino Oscillations v.s. Leptogenesis in SO(10) Models}
\author{Emmanuel Nezri \& Jean Orloff\\
Laboratoire de Physique Corpusculaire\\
F-63177 Aubi{\`e}re Cedex}

\abstract{ We study the link between neutrino oscillations and leptogenesis
  in the minimal framework assuming an \(SO(10)\) see-saw mechanism with 3
  families. Dirac neutrino masses being fixed, the solar and atmospheric
  data then generically induce a large mass-hierarchy and a small mixing
  between the lightest right-handed neutrinos, which fails to produce
  sufficient lepton asymmetry by 5 orders of magnitudes at least. This
  failure can be attenuated for a very specific value of the mixing \( \sin
  ^{2}2\theta _{e3}=0.1 \), which interestingly lies at the boundary of the
  CHOOZ exclusion region, but will be accessible to future long baseline
  experiments.  }

\preprint{LPC-RI-00-08} 
\begin{document}

\section{Introduction}

The evidence for a total asymmetry between baryon and anti-baryon densities
in our surroundings is an indisputable fact of life. This could easily be
accounted for by assuming that the initial conditions for our universe are
such that this asymmetry is of order one. However, there is every reason to
believe that the thermal history of the universe can be traced back in time
using known particle physics up to temperatures of maybe 100 GeV, certainly
high enough to massively produce quark-antiquark pairs. At these
temperatures, big-bang nucleosynthesis requires\cite{Steigman:2002qv} that
the adiabatic invariant ratio
\[
Y_{B}\doteq \frac{n_{B}-n_{\bar{B}}}{s}=0.7\, 10^{-10}\] meaning that there
be about one extra quark for \( 10^{10} \) quark-antiquark pairs. Avoiding
such extreme fine-tuning by a dynamical mechanism able to produce this
number out of an initially symmetric configuration is the purpose of
baryogenesis.

Many mechanisms of baryogenesis have been proposed in the past 30 years
(see e.g. \cite{Dolgov:1992fr} for a review including the most ingenious
and exotic ones), but most rely on ad-hoc new physics. We certainly would
prefer to explain the baryon asymmetry from solid, experimentally tested
physics. The standard model of particle physics satisfy this criterion at a
desperately high level of precision, and it was shown
\cite{Shaposhnikov:1987tw} (see \cite{Rubakov:1996vz} for a review) to
satisfy in principle all the Sakharov\cite{Sakharov:1967dj} conditions
required to produce a baryon asymmetry. However, what looked like a very
small number for initial conditions, now appears too large for the pure
standard model to achieve. At least, CP violation beyond the known CKM
phase must be added to resist the strong GIM cancellations in the hot
plasma \cite{Gavela:1994ts},\cite{Gavela:1994dt}.

Furthermore, the only possibility to change baryon number in the standard
model is via sphaleron processes which should freeze abruptly below the
electroweak phase transition to leave a net
asymmetry\cite{Bochkarev:1990fx}. Since the transition gets weaker when
rising the mass of a single scalar doublet, extra non-standard scalars are
invoked to counter the rising of the experimental lower bound on the
lightest scalar. Supersymmetry naturally provides a lot of well-motivated
extra scalars, but getting a strong enough transition with \( m_{H}\approx
100\mathrm{GeV} \) requires a huge gap between the right-handed stop and
all other sfermions, which may seem contrived, and even so, could not
protect E-W baryogenesis against a \( \approx 110\mathrm{GeV} \) Higgs
bound\cite{Carena:1997ki}. Unless a scalar is soon discovered, we thus seem
to be lead back into the pre-\cite{Kuzmin:1985mm} situation, where
baryogenesis was a footprint from extreme high-energy physics, with little
chance of an experimental cross-check.

However, another experimental signal for non-standard physics has since
then developed into a quantitative and solid field, namely the evidence for
neutrino oscillations which strongly points toward small but non-zero
neutrino masses\cite{Petcov:1999ts}.  If these masses are of the Dirac
type, right-handed neutrinos must be added, but we are left with the
puzzle: why are neutrinos \( 10^{10} \) times lighter than charged leptons?
In our mind, the smallest theoretical price to pay for resolving this
puzzle is to keep the right-handed neutrinos, but give them large Majorana
masses \( \approx 10^{10-16}\mathrm{GeV} \): this is the celebrated see-saw
mechanism \cite{Yanagida:1979xy},\cite{Gell-Mann:1980vs},
\cite{Yanagida:1980xy},\cite{Mohapatra:1980ia}.  Such a high Majorana mass
breaks lepton number at a slow enough rate, and in a very indirect way, one
can say that neutrino oscillations provide a leptonic analogue of the
baryon number violation looked for unsuccessfully in proton decay searches
during the last decades.

With this theoretical prejudice, we are thus lead to take the existence of
heavy right-handed neutrinos for granted. This makes leptogenesis
\cite{Fukugita:1986hr} an extremely natural mechanism to consider for
producing the baryon asymmetry. Indeed, the decay of right-handed neutrinos
can easily leave a CP-odd lepton asymmetry, which standard sphaleron
processes can convert into a baryon asymmetry. It is then most interesting to see how closely can this asymmetry be related to tested or testable neutrino
oscillation physics, and how much asymmetry can be produced.

To study these questions, we first review in section \ref{sec:leptogen}
the generic leptogenesis mechanism, to fix notations and give the formula
for the asymmetry in terms of neutrino masses. Then we focus in section
\ref{sec:so10} on a generic class of 3 families \( SO(10) \) models
offering a computable connection between leptogenesis and neutrino
oscillations. In section \ref{sec:ss2d}, we analytically work out this
connection in a toy see-saw model with only 2 families, to help and
understand the results presented in section \ref{sec:results}.

\section{Leptogenesis\label{sec:leptogen}}

One of the most attractive scenario to explain the baryon asymmetry of the
universe is leptogenesis. In such a scenario, a primordial lepton asymmetry
is generated by the out-of-equilibrium decay of heavy right-handed Majorana
neutrinos \( N_{i} \).  The possible excess in lepton number \( L \) is
converted into a baryon asymmetry \( B \) by the sphaleron processes. From
relations between chemical potentials this conversion is given by (see
\cite{Harvey:1990qw},\cite{Pilaftsis:1998pd} and references therein):
\begin{equation}
\label{Bchempot}
B=\frac{8N_{F}+4N_{H}}{22N_{F}+13N_{H}}(B-L)
\end{equation}
where \( N_{F} \) and \( N_{H} \) are respectively the number of families
and Higgs doublets. 

At low energies, right-handed neutrinos (\( N_{i}) \) are gauge-singlets, and
only appear in the Lagrangian via their Majorana masses (\( M_{i} \)) and Dirac
Yukawa couplings \( (h_{li}) \) to the Higgs doublet (\( H \)) and lepton
doublet (\( L_{l} \)) 
\begin{equation}
\label{Lleptons}
{\cal L}={\cal L}_{M}+{\cal L}_{Y}=\frac{1}{2}
\overline{(N_{i})^{c}}M_{i}N_{i}+
\overline{L_{l}}i\sigma _{2}H^{*}h_{li}N_{i}+h.c
\end{equation}
in the right-handed neutrinos eigenbasis \( (0<M_{1}\leq M_{2}\leq M_{3})
\). By the see-saw mechanism, light neutrinos then get a Majorana
mass-matrix :
\begin{equation}
\label{seesaw1}
m_{\nu }=v^{2}h.\diag (\frac{1}{M_{1}},\frac{1}{M_{2}},\frac{1}{M_{3}}).h^{T}
\end{equation}
The 3 Sakharov conditions are easily met by the \( N \)-decays. Indeed, the
right Majorana mass fulfils the first condition by violating lepton number
\( L \), manifest in the simultaneous existence of both decays channels \(
N_{i}\rightarrow H+l \) and \( N_{i}\rightarrow H^{\dagger }+\overline{l}
\).

Second, the CP violation condition can be met by the interference between
tree-level and \( \epsilon ' \) (vertex) or \( \epsilon \) (wave function)
one-loop corrections.  Indeed, CP asymmetry in the decay (in vacuum) of the
right-handed neutrino \( N_{i} \) is \cite{Plumacher:1998bv,Buchmuller:1998yu}:
\begin{equation}
\label{CPasym}
\delta _{i}=\frac{\sum_l \Gamma (N_{i}\rightarrow l+H)
-\Gamma (N_{i}\rightarrow \overline{l}+H^{\dagger })}
{\sum_l \Gamma (N_{i}\rightarrow l+H)
+\Gamma (N_{i}\rightarrow \overline{l}+H^{\dagger })}
=\sum_{j} \epsilon _{i,j}+\epsilon^{'} _{i,j}
\end{equation}
with 
\begin{equation}
\label{epsilon}
\epsilon _{i,j}=-\frac{1}{8\pi }
  \frac{\im (\sum_l h_{li}^{*}h_{lj})^{2}}{ \sum_l \left|h_{li}\right|^2 }
\frac{M_{i}M_{j}}{M^{2}_{j}-M^{2}_{i}}
\; ;\; 
\epsilon ^{'}_{i,j}=\frac{1}{8\pi }
  \frac{\im (\sum_l h_{li}^{*}h_{lj})^{2}}{ \sum_l \left|h_{li}\right| ^{2}}
f(\frac{M^{2}_{j}}{M^{2}_{i}})
\end{equation}
where \( f(x)=\sqrt{x}\left[ 1-(1+x)\ln (\frac{1+x}{x})\right] \).
Defining
\begin{equation}
\label{Aij}
A_{ij}=(h^{\dagger }h)_{ij}
\end{equation}
the virtual $N_j$ contribution to the $N_i$ decay asymmetry becomes for a
large hierarchy\footnote{The expressions (\ref{epsilon}) anyway require
  resumming in the degenerate limit, as the asymmetry should then vanish
  instead of
  diverging\cite{Covi:1996wh,Buchmuller:1998yu,Pilaftsis:1998pd,Frere:1999rh}.
} between heavy Majorana neutrinos \( M_{i}\ll M_{j} \)
\begin{equation}
\label{epsilon1}
\epsilon _{i,j}=-\frac{1}{8\pi }
 \frac{\im (A^{2}_{ij})}{A_{ii}}
 \frac{M_{i}}{M_{j}}\; ;\; 
\epsilon ^{'}_{i,j}=-\frac{1}{16\pi }
 \frac{\im (A^{2}_{ij})}{A_{ii}}
 \frac{M_{i}}{M_{j}}
\end{equation}
while in the opposite limit $M_i\gg M_j$:
\begin{equation}
\label{epsilon2}
\epsilon _{i,j}=-\frac{1}{8\pi }
 \frac{\im (A^{2}_{ji})}{A_{ii}}
 \frac{M_{j}}{M_{i}}\; ;\; 
\epsilon ^{'}_{i,j}=\frac{1}{4\pi }
 \frac{\im (A^{2}_{ji})}{ A_{ii} }
 \frac{M_{j}}{M_{i}}\ln (\frac{M_{j}}{M_i})\;.
\end{equation}

Concerning the third Sakharov condition, let us first assume \( N_{i} \)
to be thermally distributed in the early universe when \( T\geq M_{i} \).
While cooling, the equilibrium distribution starts feeling a strong
Boltzmann suppression when \( T\approx M_{i} \).  If the vacuum decay rate
\( \Gamma _{N_{i}}=\frac{1}{8\pi }A_{ii}M_{i} \) was much slower than the
expansion rate at that time \(
H(M_{i})=1.66\sqrt{g^{*}}\frac{M_{i}^{2}}{M_{pl}} \), each \( N_{i} \)
would decay essentially in vacuum and starting from \(
\frac{n_{i}}{n_{\gamma }}\approx 1 \) , we would get \cite{Kolb-Turner} \(
Y_{i}=L_{i}/s\approx \delta _{i}/g^{*} \).  However, decay rates are not
necessary that slow, and if the universe is still hot by the time \(
N_{i} \) decays, inverse decays dilute \( \delta _{i} \) roughly into \(
\delta_{i}/K_{i} \) with :
\[
K_{i}=\frac{\Gamma _{N_{i}}(T=M_{i})}{H(T=M_{i})}\approx \frac{1}{8\pi 
  1.66\,\sqrt{g^{*}}}A_{ii}\frac{M_{pl}}{M_{i}}
\] 
characterising the efficiency of (inverse-)decay processes when $N_i$
becomes non-relativistic. It is convenient \cite{Plumacher:1997ru} to
rewrite this dimensionless parameter in terms of an effective see-saw
like mass
\[
\tilde m_i\doteq \frac{v^2 A_{ii}}{M_{i}}=K_i \, \tilde m^*
\]
with a critical value \(\tilde m^*=\sqrt{512 g^*\pi^5/90}\;v^2/M_{pl}=1.08\ 
10^{-3}\)eV in the Standard Model. Below this critical value, decays happen
like in vacuum, but the Yukawa couplings are too small for the $N_i$
population to reach equilibrium, and the result depends on the choice of
initial conditions. If we assume no initial population, there is also a
suppression for $\tilde m_i<\tilde m^*$, and the final asymmetry
\[
Y_{i}=\frac{\delta_i}{g^*} d(\tilde m_i,M_i)
\]
contains a dilution factor $d(\tilde m)$ that is only close to 1 for
$\tilde m\approx \tilde m^*$. For $M_i\leq 10^{8}$GeV where 2 body
scatterings can be neglected, the solution of the Boltzmann equations (fig.
6 of \cite{Buchmuller:2000as}) can be reasonably fitted by
\[
\label{ourfit}
d(\tilde m)=\frac{1}{(\tilde m/10\tilde m^*)^{-0.8} + (\tilde
  m/0.28 \tilde m^*)^{1.25}}
\]
which is compared with the usual approximation\cite{Kolb-Turner}
$d(K>10)\approx 0.3(K\ln^{0.6}K)^{-1}$ on figure~\ref{dilutionfit}
(see also \cite{Buchmuller:2002rq}).

\FIGURE{
\psfrag{-7}[c][c]{\scriptsize -7}
\psfrag{-6}[c][c]{\scriptsize -6}
\psfrag{-5}[c][c]{\scriptsize -5}
\psfrag{-4}[c][c]{\scriptsize -4}
\psfrag{-3.5}[r][r]{\scriptsize -3.5}
\psfrag{-3}[c][c]{\scriptsize -3}
\psfrag{-2.5}[r][r]{\scriptsize -2.5}
\psfrag{-2}[c][c]{\scriptsize -2}
\psfrag{-1.5}[r][r]{\scriptsize -1.5}
\psfrag{-1}[c][c]{\scriptsize -1}
\psfrag{-0.5}[r][r]{\scriptsize -0.5}
\psfrag{0}[c][c]{\scriptsize 0}
\psfrag{1}[c][c]{\scriptsize 1}
\psfrag{2}[c][c]{\scriptsize 2}
\psfrag{3}[c][c]{\scriptsize 3}
\psfrag{Log[m1/eV]}[tc][tc]{$Log_{10}[\tilde m/\rm{eV}]$}
\psfrag{Log[d]}[c][r][1][90]{$Log_{10}[d]$}
\psfrag{Log[K]}[bc][bc]{$Log_{10}[K]$}
\includegraphics[width=0.75\textwidth]{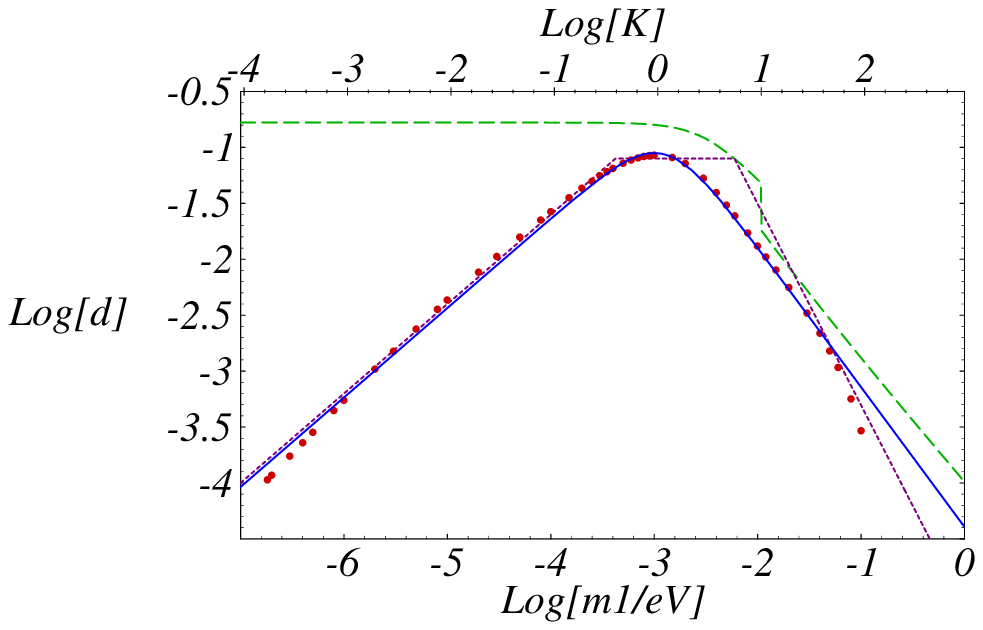}
\caption{  
  Fits of the dilution factor $d$ computed by \cite{Buchmuller:2000as} in
  the SM for $M_1=10^8$GeV (red dots): the plain blue curve is our formula
  (\ref{ourfit}), to be compared with: 1) the usual approxmation
  $0.3(K\ln^{0.6}K)^{-1}$ for large $K$ from \cite{Kolb-Turner}, extended
  to lower $K$'s following \cite{Nielsen:2001fy} (green dashed curve); 2)
  the expression of \cite{Hirsch:2001dg}, more appropriate for larger $M_1$
  (purple dotted curve).}
\label{dilutionfit}
}
 
Decay of the lightest $N_1$ usually gives the latest and dominant
contribution, because comparing (\ref{epsilon1}) and (\ref{epsilon2}) shows
that other $N_j$'s are suppressed by $A_{11}^2/A_{jj}^2${} (typically
$10^{-10}$) and further diluted by scattering processes (see for instance
\cite{Barbieri:1999ma} for a discussion of such processes).  Using \(
Y_B\approx -1/3\,Y_L \) from sphaleron conversion (\ref{Bchempot}), the
final baryon asymmetry we use in section 5 is then:
\begin{equation}
\label{YB10}
Y_{B10}\doteq 10^{10}\,Y_B\approx 
-\frac{10^{10}}{3g^\star}d(\tilde m_1,M_1)
\frac{-3 M_1}{16\pi A_{11}} 
\sum_{j=2,3}\frac{\im (A^{2}_{1j})}{M_{j}}
\approx 0.7
\end{equation}

At this level, no constraints from neutrino oscillations encoded in
(\ref{seesaw1}) can be drawn because the Yukawa coefficients \(h\) are free
parameters: for any set of parameters \( (h,M) \) providing good
oscillation physics but bad \( Y_{B10} \), we can rescale \( h\rightarrow
h^{'}=h/\sqrt{Y_{B10}} \) and \( M\rightarrow M^{'}=M/Y_{B10} \) to get the
same light neutrinos, with \( Y_{B10}\approx 1 \). Notice that $K$ and
$\tilde m$ are invariant under such rescaling, so that the only violation
to this scaling law comes from the $M_1$ dependence of the dilution factor
$d$. 
 
To set the $M_1$ scale of interest, it is instructive to make the further
approximation $d(1<K<10^3)\approx0.1/K$ and assume domination of the $N_1$
decay with a virtual $N_2$ exchange. The result for the asymmetry is then
\begin{equation}
\label{YB10approx}
Y_{B10}\approx\frac{0.7\times 10^{-10}}{\sqrt{g^{*}}}
 \frac{\im (A^{2}_{12})}{A_{11}^{2}}.
 \frac{M_{1}}{M_{2}}.
 \frac{M_{1}}{10^{10}{\rm GeV}}\approx 0.7
\end{equation}
so that \( M_{1}\approx 10^{10}\mathrm{GeV} \) is the natural scale for
leptogenesis.  

To draw any link between neutrino oscillations and leptogenesis, we thus
need further theoretical assumptions, of which we now describe our minimal
set.

\section{SO(10)\label{sec:so10}}

\( SO(10)\supset SU(3)\times SU(2)\times U(1) \)

The \( SO(10) \) Grand Unification group is a very natural framework for
studying leptogenesis. Indeed \cite{Fritzsch:1975nn,Mohapatra}, it is the
smallest GUT group including right-handed neutrinos together with usual
fermions in matter multiplets :
\[
\underbrace{16}_{\psi }=
\underbrace{(3,2)}_{Q}+
\underbrace{(\overline{3},1)}_{u_{R}^{c}}+
\underbrace{(\overline{3},1)}_{d_{R}^{c}}+
\underbrace{(1,2)}_{L}+
\underbrace{(1,1)}_{e_{R}^{c}}+
\underbrace{(1,1)}_{N}
\]
Furthermore, it includes \( (B-L) \) as a gauge symmetry, whose breaking
gives a natural explanation for right-handed neutrino Majorana masses.
Finally, leptons and quarks masses are related which fixes the neutrino
Dirac Yukawa \( h \).  For instance, if Dirac masses originate from a
single scalar in the vector (10-dimensional) of \( SO(10) \) :
\begin{equation}
\label{Lmass10}
{\cal L}_{Y}=f^{10}_{ab}\psi _{a}^{T}\Gamma _{A}\psi _{b}\phi ^{10}_{A}
\end{equation}
 we obtain the (unphysical) low energy relation :
\[
m_{d}=3m_{e}=m_{u}=3m_{\nu }\propto f^{10}v^{10}\]
where the factor of about 3 comes from the extra color loops affecting quarks
but not leptons. Adding a second (or more) scalars in the same \( SO(10) \)
representation :
\begin{equation}
\label{Lmass10'}
{\cal L}^{'}_{Y}=f^{'10}_{ab}\psi _{a}^{T}\Gamma _{A}\psi _{b}\phi ^{'10}_{A}
\end{equation}
lifts the degeneracy \( m_{u}=m_{d} \) while keeping \( m_{u}\approx
3m_{\nu } \) i.e \( m_{\nu }\approx 1{\rm MeV}\rightarrow 50{\rm GeV} \).

To get more realistic neutrino masses, an extra scalar \( \phi ^{126} \) in
the 126-representation with v.e.v \( v_{126} \) in the \( SU(5) \)-singlet
component can be invoked to break \( (B-L) \), providing the Majorana mass
term : 
\begin{equation}
\label{Lmass126}
{\cal L}_{M}=f_{ab}^{126}v_{126}\overline{N_{a}^{c}}N_{b}+h.c
\end{equation}
Thus, in the framework of \( SO(10) \), Dirac masses are naturally determined
by up-quarks masses.

\subsubsection*{\protect\( A_{ij}\protect \) expression}

Writing the relevant Lagrangian term for the leptons we get after \( SU(2) \)
breaking :
\begin{equation}
\label{Lleptons2}
{\cal L}=\left[ \overline{L_{a}}\times 
\left( \begin{array}{c} 0\\ v^{d} \end{array}\right)
 \right] 
 f^{d}_{ab}l_{Rb}+\left[ \overline{L_{a}}\times 
\left( \begin{array}{c} v^{u}\\ 0 \end{array}\right)
 \right] 
 f^{u}_{ab}N_{b}+\frac{1}{2}\overline{N_{a}^{c}}v_{126}f^{126}_{ab}N_{b}+h.c
\end{equation}
so that \( M_{ab}\doteq f^{126}_{ab}v_{126} \) with \( a \) and \( b \) being
components in some arbitrary family basis. Decomposing the Yukawa couplings
into mixing and real masses eigenvalues, we write :

\begin{eqnarray}
f^{d} & = & \frac{1}{3v^{d}}
  U_{Ld}^{\dagger}.\diag (m_{d},m_{s},m_{b}).U_{Rd}\label{yukd} \\
f^{u} & = & \frac{1}{3v^{u}}
  U_{Lu}^{\dagger}.\diag (m_{u},m_{c},m_{t}).U_{Ru}\label{yukup} \\
f^{126} & = & \frac{1}{v^{126}}
  U_{M}^{*}.\diag (M_{1}<M_{2}<M_{3}).U^{\dagger }_{M}\label{yukMajo} 
\end{eqnarray}
The previous \( h_{li} \) are up-Yukawa matrix elements between charged
leptons mass (and flavour) eigenstates \( (l_{l}) \) and right-handed
neutrinos mass eigenstates \( (N_{i}) \). Therefore:

\begin{eqnarray}
h_{li} & = & U_{Ld}.f^{u}.U_{M}\\
& = & \frac{1}{3v^{u}}
 U_{Ld}.U^{\dagger }_{Lu}.\diag (m_{u},m_{c},m_{t}).U_{Ru}.U_{M}\label{yukli}\\
& \doteq  & \frac{1}{3v^{u}}
 V^{\dagger }_{CKM}.\diag (m_{u},m_{c},m_{t}).V_{R}
\end{eqnarray}
so that
\begin{equation}
\label{Aij2}
A_{ij}=\frac{1}{9(v^{u})^{2}}
V^{\dagger }_{R}.\diag (m^{2}_{u},m^{2}_{c},m^{2}_{t}).V_{R}
\end{equation}
only depend on the unknown \( V_{R} \) mixing. But the light neutrinos mass
matrix is constrained by neutrinos oscillation physics:
\begin{eqnarray}
m_{\nu } & = & v_{u}^{2}h.
\diag(\frac{1}{M_{1}},\frac{1}{M_{2}},\frac{1}{M_{3}}).h^{T}\label{mnuseesaw}\\
 & = & U.\diag (m_{1},m_{2},m_{3}).U^{T}\label{mnuoscill} 
\end{eqnarray}
The unknowns \( (V_{R},M_{1,2,3}) \) can thus be determined from measurable
quantities \( (U,m_{1,2,3}) \) by diagonalizing the matrix :
\begin{eqnarray}
M^*_{R} & = & \frac{1}{9}
\diag (m_{u},m_{c},m_{t}).U_{eff}.
\diag (\frac{1}{m_{1}},\frac{1}{m_{2}},\frac{1}{m_{3}}).U^{T}_{eff}.
\diag (m_{u},m_{c},m_{t})\label{MRseesaw} \\
 & = & V_{R}.\diag (M_{1},M_{2},M_{3}).V^{T}_{R}\label{MRdiag} 
\end{eqnarray}
where
\begin{equation}
\label{Ueff}
U_{eff}\doteq V^{CKM}.U
\end{equation}

\subsubsection*{Parameterisation \& experimental inputs: }

Using the standard parameterisation\cite{PDG02} for \( V^{CKM} \) :
\begin{equation}
V^{CKM}=V_{23}.V_{13}.V_{12}=\left( \begin{array}{ccc}
c_{12}c_{13} & s_{12}c_{13} & s_{13}^{*}\\
-s_{12}c_{23}-c_{12}s_{23}s_{13} & c_{12}c_{23}-s_{12}s_{23}s_{13} & s_{23}c_{13}\\
s_{12}s_{23}-c_{12}c_{23}s_{13} & -c_{12}s_{23}-s_{12}c_{23}s_{13} & c_{23}c_{13}
\end{array}\right) 
\label{VCKM}
\end{equation}
 we use the following values for the quark mixing angles:
\begin{eqnarray*}
s_{23}^{CKM} & = & 0.041\\
s_{13}^{CKM} & = & 0.0036\,e^{i\delta^{CKM}}\,;\,\delta^{CKM}\approx 1\\
s_{12}^{CKM} & = & 0.223
\end{eqnarray*}

As for light neutrinos\cite{Bilenkii:1998dt}, we label mass eigenstates so that
the most constraining experiments each fix a definite mixing angle and use a
similar decomposition with three complex mixing angles, and Majorana phases:
\begin{eqnarray}
\label{Uneutrinos}
U & =
&\underbrace{U_{23}}_{atm}.
 \underbrace{U_{13}}_{Chooz}.
 \underbrace{U_{12}}_{sun}.
 \diag(1,e^{i\phi_1},e^{i\phi_2})\\
&=&
\left( \begin{array}{ccc}
c_{sun}c_{e3} & s_{12}^{*}c_{e3} & s_{13}^{*}\\
-s_{12}c_{atm}-c_{sun}s_{23}^{*}s_{13} & 
  c_{sun}c_{atm}-s_{12}^{*}s_{23}^{*}s_{13} & 
  s_{23}^{*}c_{e3}\\
s_{12}s_{23}-c_{sun}c_{atm}s_{13} & 
  -c_{sun}s_{23}-s_{12}^{*}c_{atm}s_{13} & 
  c_{atm}c_{e3}
\end{array}\right) .
\left( \begin{array}{ccc}
1 & 0 &0\\
0 & e^{i\phi_1} & 0\\
0 & 0 &  e^{i\phi_2}
\end{array}\right)\nonumber
\end{eqnarray}
where the complex parameters are:
\begin{eqnarray*}
  s_{12}&=&s_{sun}e^{i\delta_{12}}\\
  s_{13}&=&s_{e3}e^{i\delta_{13}}\\
  s_{23}&=&s_{atm}e^{i\delta_{23}}\\
\end{eqnarray*}
 Squared mass differences are then fixed (up to a sign) by the oscillation parameters
\( \Delta m^{2}_{sun}\ll \Delta m^{2}_{atm} \):
\begin{eqnarray}
m^{2}_{2}-m^{2}_{1} & = & \Delta m^{2}_{sun}\label{deltam2sun} \\
m^{2}_{3}-m^{2}_{1} & \approx  & m^{2}_{3}-m^{2}_{2}
 = \pm \Delta m^{2}_{atm}\label{deltam2atm} 
\end{eqnarray}
In both cases, \( m_{3} \) is singled out by atmospheric oscillations, but
according to the sign, it can either be heavier
\[
m_{1}\le m_{2}\ll m_{3}\qquad (\mathrm{standard}\, \mathrm{hierarchy}),\]
 or lighter than the others

\[
m_{3}\ll m_{1}\le m_{2}\qquad (\mathrm{inverted}\, \mathrm{hierarchy}).\] We
take the lightest mass (\( m_{1} \) or \( m_{3} \)) as the free parameter that
cannot be fixed by oscillations. For values of this parameter larger than \(
\Delta m_{atm} \), we get two different realizations of the degenerate case.
In the following, it will vary between \( 10^{-7}\mathrm{eV} \) and \(
1\mathrm{eV} \): going below this range requires huge right-handed masses,
typically above GUT or even Planck scale; going above does not help to increase
the asymmetry, and generally violates neutrino-less double beta decay
experiments\cite{Baudis:1999xd,Giunti:1999jw} (see however
\cite{Vissani:1999tu} for a study of possible cancellations using Majorana
phases).

Other free parameters are the six phases: $\delta^{CKM}$ (from $V^{CKM}$),
$\delta_{12}$, $\delta_{13}$, $\delta_{23}$ (from the complex mixing
angles\footnote{Notice that we take $\delta_{12,13,23}\in [-\pi/2,\pi/2]$,
  so that $s_{sun,e3,atm}\in[-1,1]$ and $c_{sun,e3,atm} \in[0,1]$ instead
  of being restricted to the first quadrant like the usual CKM angles} in
$U$), and the two Majorana phases $\phi_{1,2}$. A careful count of the 6
phases in see-saw models can be found in reference \cite{Endoh:2000hc}.

With the above parametrisation, the experimental constraints from oscillations
lead us to take the following values in the subsequent numerical analysis:
\begin{itemize}
\item Atmospheric neutrinos\cite{PDG02}: \( \Delta m_{atm}^{2}\approx 3\,
  10^{-3}\mathrm{eV}^{2} \) with nearly maximal mixing: \( \sin ^{2}2\theta
  _{atm}=0.92\rightarrow 1 \) so that \( s_{atm}=0.6 \rightarrow 0.7\).
\item CHOOZ reactor experiment\cite{Apollonio:1999ae}: for the above value
  of \( \Delta m_{atm}^{2} \), \( |U_{e3}|\doteq s_{e3}<0.18 \) at 95\% C.L. 
\item Solar neutrinos\cite{Gonzalez-Garcia:2002dz}: we will consider 4
  possible resolutions of the solar neutrino deficit:
\begin{enumerate}
\item large mixing angle MSW oscillations (LMA) with \( \Delta
  m^{2}_{sun}\approx 5\,10^{-5}\mathrm{eV}^{2} \) and \(
  \tan^2\theta_{sun}=0.42\in[0.15,0.9]\) at 95\% C.L. corresponding to 
  \(s_{sun}=0.54 \in[0.36,0.69]\) ;
\item small mixing angle MSW oscillations (SMA) with \( \Delta
  m^{2}_{sun}\approx 5\; 10^{-6}\mathrm{eV}^{2} \) and \(
  \tan^2\theta_{sun}=1.5\; 10^{-3}\in[6\;10^{-4},3\;10^{-3}]\) at 95\% C.L. 
  corresponding to \(s_{sun}=0.039\in[0.025,0.055] \) ;
\item LOW with \( \Delta m^{2}_{sun}\approx 7.9\; 10^{-8}\mathrm{eV}^{2} \)
  and \( \tan^2\theta_{sun}=0.61\) corresponding to \(s_{sun}=0.62 \) ;
\item vacuum oscillations (VAC) with \( \Delta m^{2}_{sun}\approx
  4.6\; 10^{-10}\mathrm{eV}^{2} \) and \( \tan ^{2}\theta _{sun}=1.8 \) so that
  \( s_{sun}= 0.8 \);
\end{enumerate}
Although the LMA solution is favoured by the SNO data \cite{Ahmad:2002ka}
and confirmed by the latest KamLAND results\cite{Eguchi:2002dm}, we still
consider the SMA, LOW and VAC situations for completeness and illustration.
\end{itemize}

\section{Toy see-saw (2 flavours)\label{sec:ss2d}}

Before reviewing the results in a full case with 3 families, it helps
intuition to work out a 2 families exercise. In particular, equation
(\ref{YB10}) suggests that leptogenesis prefers right handed
neutrinos with large mixing and similar masses. This is not easy to achieve
by see-saw with large Dirac hierarchies and
mixing\cite{Yanagida:1980ve,Akhmedov:1999ws}.  Starting from the see-saw
formula (\ref{MRseesaw}), dropping the first family and introducing the
Dirac hierarchy parameter \( r\doteq m_{c}/m_{t}\approx 10^{-2} \) for
instance and the light neutrinos hierarchy parameter \( m\doteq m_{2}/m_{3}
\), we get in the case of real effective mixing parametrised by \( s\doteq
\sin \theta _{23}^{eff} \)
\begin{eqnarray}
M_{R} & = & \frac{1}{9}\left( \begin{array}{cc}
m_{c} & 0\\
0 & m_{t}
\end{array}\right) \cdot \left( \begin{array}{cc}
c & s\\
-s & c
\end{array}\right) \cdot \left( \begin{array}{cc}
\frac{1}{m_{2}} & 0\\
0 & \frac{1}{m_{3}}
\end{array}\right) \cdot \left( \begin{array}{cc}
c & -s\\
s & c
\end{array}\right) \cdot \left( \begin{array}{cc}
m_{c} & 0\\
0 & m_{t}
\end{array}\right) \\
 & = & \frac{m_{t}^{2}}{9m_{3}}\cdot \left( \begin{array}{cc}
r^{2}(c^{2}+s^{2}m) & rcs(m-1)\\
rcs(m-1) & c^{2}m+s^{2}
\end{array}\right) \cdot \frac{1}{m}\label{MR2dgen} 
\end{eqnarray}

For a large inverse hierarchy ({\it w.r.t. } Dirac hierarchy at small
angle) \( m\gg 1 \), we get for small \( r,s\ll 1 \) (in units of \(
m^{2}_{t}/9m_{3} \))
\begin{equation}
\label{MR2dappr}
M_{R}\approx \left( \begin{array}{cc}
r^{2}s^{2}+r^{2}/m & rs\\
rs & 1
\end{array}\right) =\left( \begin{array}{cc}
c_{R} & -s_{R}\\
s_{R} & c_{R}
\end{array}\right) \cdot \left( \begin{array}{cc}
M_{1} & 0\\
0 & M_{2}
\end{array}\right) \cdot \left( \begin{array}{cc}
c_{R} & s_{R}\\
-s_{R} & c_{R}
\end{array}\right) 
\end{equation}
so that \( M_{2}\approx \Tr M_{R}\approx 1, \) \( M_{1}\approx \Det
M_{R}=r^{2}/m\ll \Tr M_{R} \) while the right-handed mixing \(
s_{R}^{2}\approx r^{2}s^{2} \) is strongly suppressed by the Dirac
hierarchy \( r\ll 1 \). The ratio appearing in the leptogenesis
(\ref{YB10}) is
\[
\frac{A_{12}}{A_{11}}=-\frac{c_{R}s_{R}}{r^{2}c_{R}^{2}+s_{R}^{2}}\approx
-\frac{s}{r}\] 
and the the asymmetry (\ref{YB10approx}) goes in this case like
\[
Y^{appr}_{B10}\propto \frac{A_{12}^{2}}{A_{11}^{2}}\cdot
\frac{M_{1}}{M_{2}}\cdot M_{1}\approx \frac{r^{2}s^{2}}{m^{2}}
\]
which is strongly suppressed and dominated by the smallest values of \( m
\) possible. But in this degenerate region \( m\approx 1 \), another type
of suppression arises as \( s_{R} \) vanishes together with the
off-diagonal terms in (\ref{MR2dgen}).

\FIGURE{
\psfrag{-6}[c][c]{-6}
\psfrag{-5}[c][c]{-5}
\psfrag{-4}[c][c]{-4}
\psfrag{-3}[c][c]{-3}
\psfrag{-2}[c][c]{-2}
\psfrag{-1}[c][c]{-1}
\psfrag{0}[c][c]{0}
\psfrag{m}[c][cb]{$Log_{10}[m]$}
\psfrag{-4}[r][r]{-4}
\psfrag{-2}[r][r]{-2}
\psfrag{0}[r][r]{0}
\psfrag{2}[r][r]{2}
\psfrag{4}[r][r]{4}
\psfrag{MR}[c][rl][1][90]{$Log_{10}[M_R]$}
\psfrag{s=0.1 >> r}[l][l]{$s=0.1 \gg r$}
\psfrag{s=0.01 = r}[l][l]{$s=0.01 = r$}
\psfrag{s=0.001 << r}[l][l]{$s=0.001 \ll r$}
\includegraphics[width=0.75\textwidth]{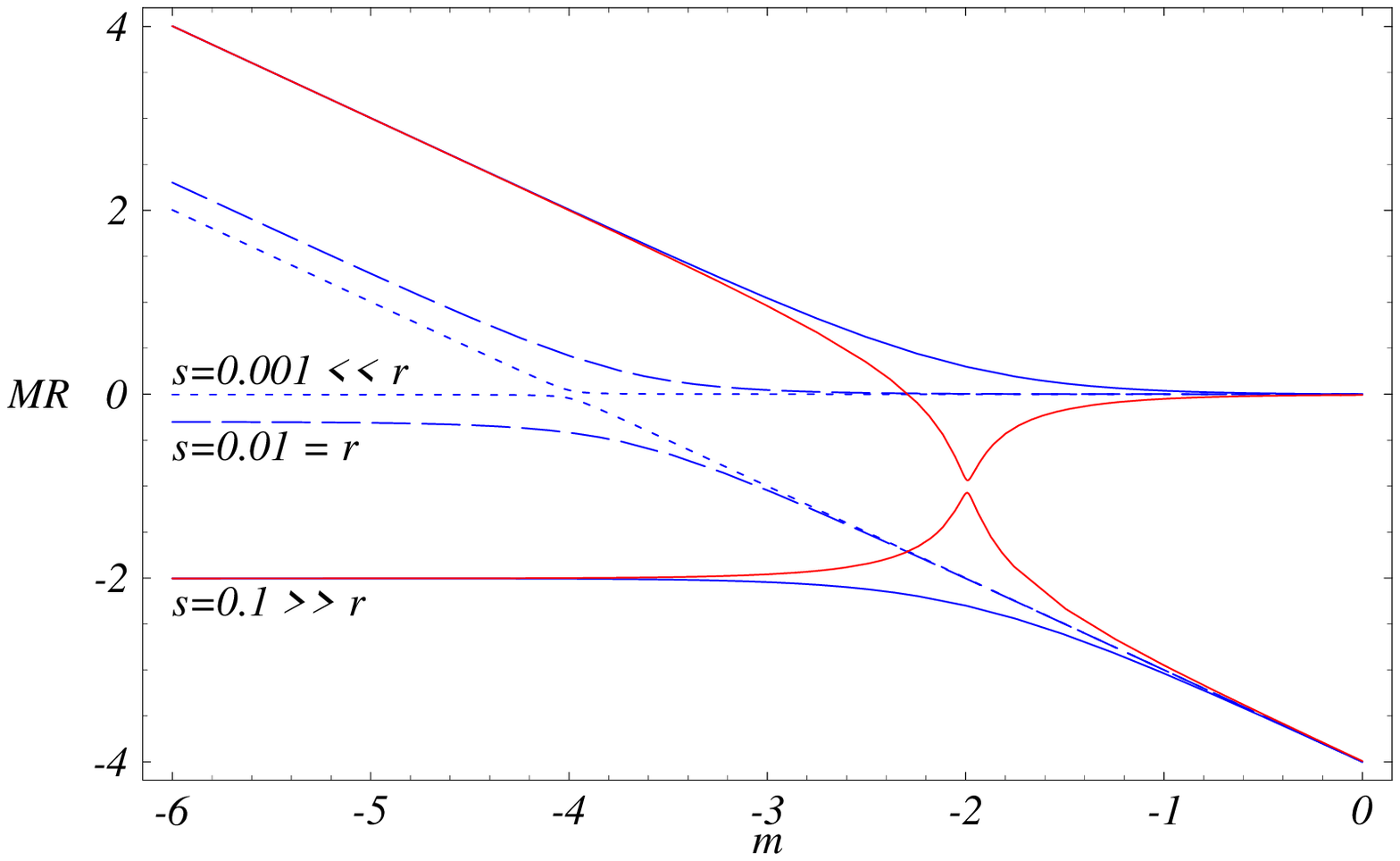}
\caption{\label{fig:ss2d-mass}
  Right-handed masses \protect\( M_{R}\protect \) in arbitrary units, as a
  functions of light neutrino masses ratio \protect\( m=m_{2}/m_{3}\protect
  \), for 3 values of light neutrino effective mixing \protect\( s=\sin
  (\theta _{23eff})=10^{-1},10^{-2},10^{-3}\protect \) and a fixed value of
  the Dirac masses hierarchy \protect\( r=m_{c}/m_{t}=0.01\protect \). For $
  s=0.1$, we also show the effect of the Majorana phase $\phi_m=0.99\pi$ (red
  curve) leading to degenerate Majorana masses.}
}

To hope for larger results, we must thus turn to the usual hierarchy limit
\( m\ll 1 \).  Assuming again \( r,s\ll 1 \), \( M_{R} \) in
(\ref{MR2dgen}) then becomes
\[
\left( \begin{array}{cc}
r^{2} & -rs\\
-rs & s^{2}+m
\end{array}\right) \cdot \frac{1}{m}\]
By furthermore assuming \( M_{1}\ll M_{2} \) (i.e. \( r^{2}m\ll
(m+r^{2}+s^{2})^{2} \) which turns out to be satisfied everywhere, except
when \( s^{2}\ll r^{2}\approx m \) (see figure \ref{fig:ss2d-mass}), we
arrive at the useful analytic expressions

\begin{eqnarray}
M_{2} &\approx & \Tr M_{R}\approx \frac{1}{m}(m+r^{2}+s^{2})\label{M2appr} \\
M_{1} &\approx & \frac{\Det M_{R}}{\Tr M_{R}}
       \approx \frac{r^{2}}{m+r^{2}+s^{2}}\label{M1appr} \\
-A_{21} & = & s_{R}c_{R}\approx \frac{-rs}{m+r^{2}+s^{2}}\label{A21appr} 
\end{eqnarray}
From these expressions, it is clear that a cross-over occurs at values of
\( m \) around the largest of \( s^{2} \) or \( r^{2} \), but unlike the
level-repulsion effect usual in quantum mechanics, the ``levels'' \(
M_{1,2} \) do not need to be close for the cross-over to happen, as can be
seen in the \( s>r \) situation.  In this case, the levels do not
asymptotically tend to the no-mixing situation away from cross-over region.
These results are most simply summarised in the plots \ref{fig:ss2d-mass},
drawn without approximation for \( r=10^{-2} \).

Since \( A_{11}=r^{2}c_{R}^{2}+s_{R}^{2} \), the ratio \( A_{12}/A_{11} \)
only takes a simple form for small mixing \( s_{R}\ll r \) (large enough \(
m \) or small enough \( s \)), in which case the expression coming into the
baryon asymmetry is simply
\begin{equation}
\label{YBapprox}
Y^{appr}_{B10}\propto \frac{A_{12}^{2}}{A_{11}^{2}}\cdot 
\frac{M_{1}}{M_{2}}\cdot M_{1}\approx \frac{ms^{2}r^{2}}{(m+r^{2}+s^{2})^{5}},
\end{equation}
growing like \( m\propto M_{2}^{-1} \) for small \( m \), reaching a
maximum around \( r^{2}+s^{2} \) and decreasing like \( m^{-4} \) beyond.
We see in figure \ref{ss2d-YB} that these approximations capture the
general shape of the asymmetry, except for extra suppressions in the
regions \( m\approx 1 \) (left-handed degenerate ), and \( m<s \) if \(
s^{2}\ll r^{2} \) (\( A_{11}\gg r^{2} \)), where the approximations used
break down.

\FIGURE{
\psfrag{-6}[c][c]{-6}
\psfrag{-5}[c][c]{-5}
\psfrag{-4}[c][c]{-4}
\psfrag{-3}[c][c]{-3}
\psfrag{-2}[c][c]{-2}
\psfrag{-1}[c][c]{-1}
\psfrag{0}[c][c]{0}
\psfrag{Log10[m]}[c][c]{$Log_{10}[m]$}
\psfrag{-6}[r][r]{-6}
\psfrag{-4}[r][r]{-4}
\psfrag{-2}[r][r]{-2}
\psfrag{0}[r][r]{0}
\psfrag{2}[r][r]{2}
\psfrag{Log[YB10]}[c][c][1][90]{$Log_{10}[Y_{B10}]$}
\includegraphics[width=0.75\textwidth]{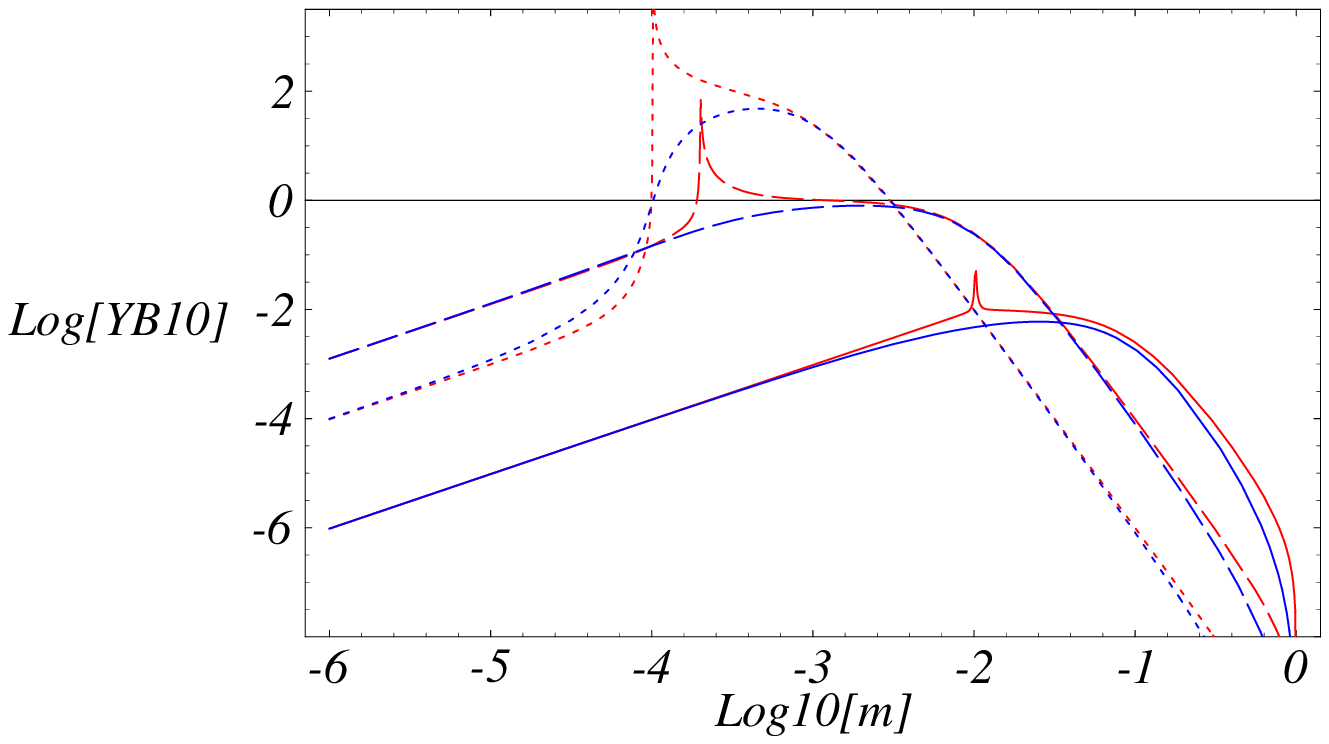}
\caption{\label{ss2d-YB}Log-Log plot of the asymmetry 
  \protect\( Y_{B10}\protect \) (in arbitrary units) as a function of
  \protect\( m=m_{2}/m_{3}\protect \) with the same conventions as in
  figure \ref{fig:ss2d-mass}. For each curve there are 2 twins: the red
  spiky curve is the exact version of (\ref{YBcorrect}) computed for the phase
  maximising its value, while the smooth blue curve is the exact version of
  (\ref{YBapprox}) for real positive parameters.}
}

Of course, the true asymmetry \( Y_{B10} \) vanishes unless \( A_{12} \)
gets a CP-violating non-zero phase. Specialising (\ref{Uneutrinos}) for 2
flavours leaves 2 phases, \(\phi_m\doteq 2(\phi_2-\phi_1)\) and
$\delta_{23}$, which respectively complexify the parameters
$m=|m|e^{i\phi_m}$, $s=|s|e^{i\delta_{23}}$ and $s_R=|s_R|e^{i\delta_R}$ in
the previous discussion. It is a bit tedious to check that the above
approximations (\ref{M2appr}-\ref{A21appr}) for \( M_{1}\ll M_{2} \)
and \( |A_{12}| \) can be extended to complex parameters, and that the
relevant phase is given by
\begin{equation}
\label{A21phase}
A_{12}=e^{-i\delta_{R}}\, |A_{12}|.
\end{equation}
The asymmetry (\ref{YB10approx}) in the same approximation as (\ref{YBapprox}) is then
actually
\begin{equation}
\label{YBcorrect}
Y_{B10}\propto \frac{\im (A_{12}^{2})}{A_{11}^{2}}\cdot 
\frac{M_{1}}{M_{2}}\cdot M_{1}\approx \sin (-2\delta_{R})
\frac{|m|r^{2}|s|^{2}}{|m+r^{2}+s^{2}|^{5}}
\end{equation}
which has the same qualitative features as the previous expression \(
Y_{B10}^{appr} \), with a further suppression by the factor \(
|\sin(-2\delta_{R})|<1 \), and a possible enhancement when the
denominator cancels, {\it i.e.} outside the range of validity of the
approximations. Figure \ref{ss2d-YB} shows that for a maximising value of
the phase \(\delta_{R}\), the correct asymmetry \(
Y_{B10} \) is close to \( Y_{B10}^{appr} \) worked out without phases,
after replacing \( \im (A_{12}^{2}) \) by \( |A_{12}^{2}|=A_{12}^{2} \). This
approximation can be a useful guide for finding upper bounds in the full 3
flavours case, where the large number of phases makes it non-trivial.

To summarise, the lesson learned in this 2 flavours see-saw exercise is that 

\begin{enumerate}
\item the right neutrinos mass ratio \( M_{1}/M_{2} \) is maximised for a
  ratio of light neutrino masses \( m=m_{1}/m_{2} \) which is either the
  squared ratio of Dirac masses \( r^{2}=(m_{D1}/m_{D2})^{2} \) or the
  squared sine of the light effective mixing angle \( s^{2}=\sin ^{2}\theta
  _{eff} \), whichever is largest,
\item the baryon asymmetry can be maximised for a similar or slightly
  larger value of \( m \),
\item Majorana phases cancellations can lead to further enhancements around
  that same value of $m$ (red curve on figure \ref{fig:ss2d-mass}), but these
  can only be trusted after a careful resummation of this degenerate
  $M_1\approx M_2$ case;
\item for small mixing, the ``inverted hierarchy'' \( m_{1}\gg m_{2} \)
  gives much smaller results than the ``standard hierarchy'' \( m_{1}\ll
  m_{2} \).
\end{enumerate}

Finally, we should also notice that this 2 flavours exercise was
based on the approximate form (\ref{YBapprox}) which, as
(\ref{YB10approx}), is only valid for $K>1$, and overestimates the
asymmetry for lower $K$. For the results below we use the
relation (\ref{YB10}) which includes more realistic Boltzmann dilution.

\section{Results with 3 flavours \label{sec:results}}

Let us now discuss the more realistic situation of 3 light neutrino
flavours.  It turns out the baryon asymmetry is generically too small to be
useful. Before concluding that \( SO(10) \) see-saw leptogenesis is
excluded, we must look for regions of parameters where the asymmetry can be
maximised. As seen in the previous section, standard hierarchy is a better
candidate in this respect.  We will pass in review the various solutions to
the solar neutrino deficit, starting by vacuum oscillations (VAC).

\FIGURE{
\psfrag{-7}[c][c]{-7}
\psfrag{-6}[c][c]{-6}
\psfrag{-5}[c][c]{-5}
\psfrag{-4}[c][c]{-4}
\psfrag{-3}[c][c]{-3}
\psfrag{-2}[c][c]{-2}
\psfrag{-1}[c][c]{-1}
\psfrag{0}[c][c]{0}
\psfrag{m1}[ct][cb]{$Log_{10} m_1$}
\psfrag{4}[r][r]{4}
\psfrag{6}[r][r]{6}
\psfrag{8}[r][r]{8}
\psfrag{10}[r][r]{10}
\psfrag{12}[r][r]{12}
\psfrag{14}[r][r]{14}
\psfrag{16}[r][r]{16}
\psfrag{MR}[c][l][1][90]{$Log_{10} [M_R]$}
\psfrag{VAC: Right handed masses}[c][c]{Vacuum: Right handed masses}
\psfrag{se3=0}[tl][lb]{{\darkblue$s_{e3}=0$}}
\psfrag{se3=-0.16}[tl][lb]{{\darkred$s_{e3}=-0.16$}}
\psfrag{Ueff=1}[l][lt]{{\darkgreen$U_{eff}=1$}}
\psfrag{M1}[r][r]{$M_1$}
\psfrag{M2}[r][r]{$M_2$}
\psfrag{M3}[r][r]{$M_3$}
\includegraphics[width=0.75\textwidth]{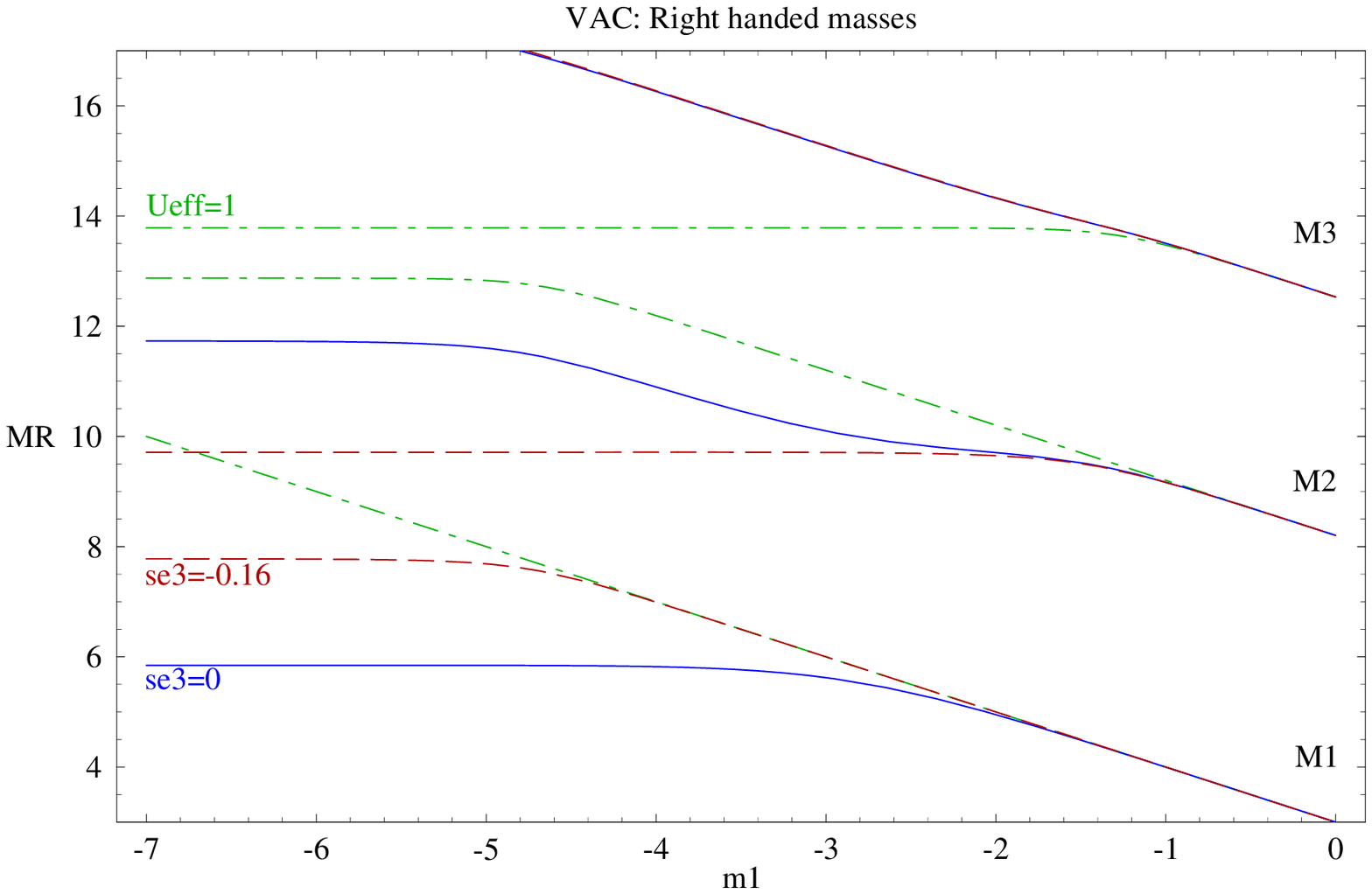}
\caption{\label{3MRVO}Masses of right-handed heavy Majorana neutrinos 
  for vacuum oscillation of solar neutrinos (\protect\( \Delta
  m_{sun}=2.15\,10^{-5}\mathrm{eV}\protect \)). The dash-dotted curves completely
  ignore mixing (\protect\( U_{eff}=1\protect \)), the plain curves are for
  real CKM mixing, maximal atmospheric and solar mixing but vanishing
  \protect\( s_{e3}\protect \) while the dashed curves tune \protect\(
  s_{e3}\protect \) to maximise \protect\( M_{1}/M_{2}\protect \).}
}

In view of the previous section, a good starting point to maximise the
baryon asymmetry is to try and make the lightest right-handed masses \(
M_{1} \) and \( M_{2} \) as close as possible to each other. If there were
no mixing, \( U_{eff} \) in (\ref{MRseesaw}) would be diagonal and
right-handed masses would follow the dash-dotted curves of
figure~\ref{3MRVO}. For large \( m_{1} \) (degenerate limit) they draw 3
parallel lines separated by large Dirac mass hierarchies.  With decreasing
\( m_{1} \), all \( M_{R} \)'s grow until \( m_{1}=\Delta m_{atm}\approx
10^{-1}\mathrm{eV} \), below which \( M_{3} \) levels off at \(
m_{t}^{2}/9\Delta m_{atm}\approx 10^{14}\mathrm{GeV} \).  Meanwhile \(
M_{1,2} \) continue growing together until \( m_{1}=\Delta m_{sun}\approx
10^{-5}\mathrm{eV} \), where \( M_{2} \) stops at about the same value.
Further decreasing \( m_{1} \) then only affects \( M_{1} \), which becomes
degenerate with \( M_{2} \) for extremely low \( m_{1}\approx
10^{-11}\mathrm{eV} \).

\FIGURE{
\psfrag{-7}[c][c]{-7}
\psfrag{-6}[c][c]{-6}
\psfrag{-5}[c][c]{-5}
\psfrag{-4}[c][c]{-4}
\psfrag{-3}[c][c]{-3}
\psfrag{-2}[c][c]{-2}
\psfrag{-1}[c][c]{-1}
\psfrag{0}[c][c]{0}
\psfrag{Log[m1/eV]}[ct][cb]{$Log_{10}[m_1/{\rm eV}]$}
\psfrag{-8}[r][r]{-8}
\psfrag{-6}[r][r]{-6}
\psfrag{-4}[r][r]{-4}
\psfrag{-2}[r][r]{-2}
\psfrag{0}[r][r]{0}
\psfrag{Log[YB10]}[bc][bc][1][90]{$Log_{10}[Y_{B10}]$}
\psfrag{VAC: se3=0}[l][l]{\darkblue VAC: $s_{e3}=0$}
\psfrag{VAC: se3=-0.16}[l][l]{\darkred VAC: $s_{e3}=-0.16$}
\psfrag{LOW: se3=-0.16}[l][bl]{{\darkgreen LOW: $s_{e3}=-0.16$}}
\includegraphics[width=0.75\textwidth]{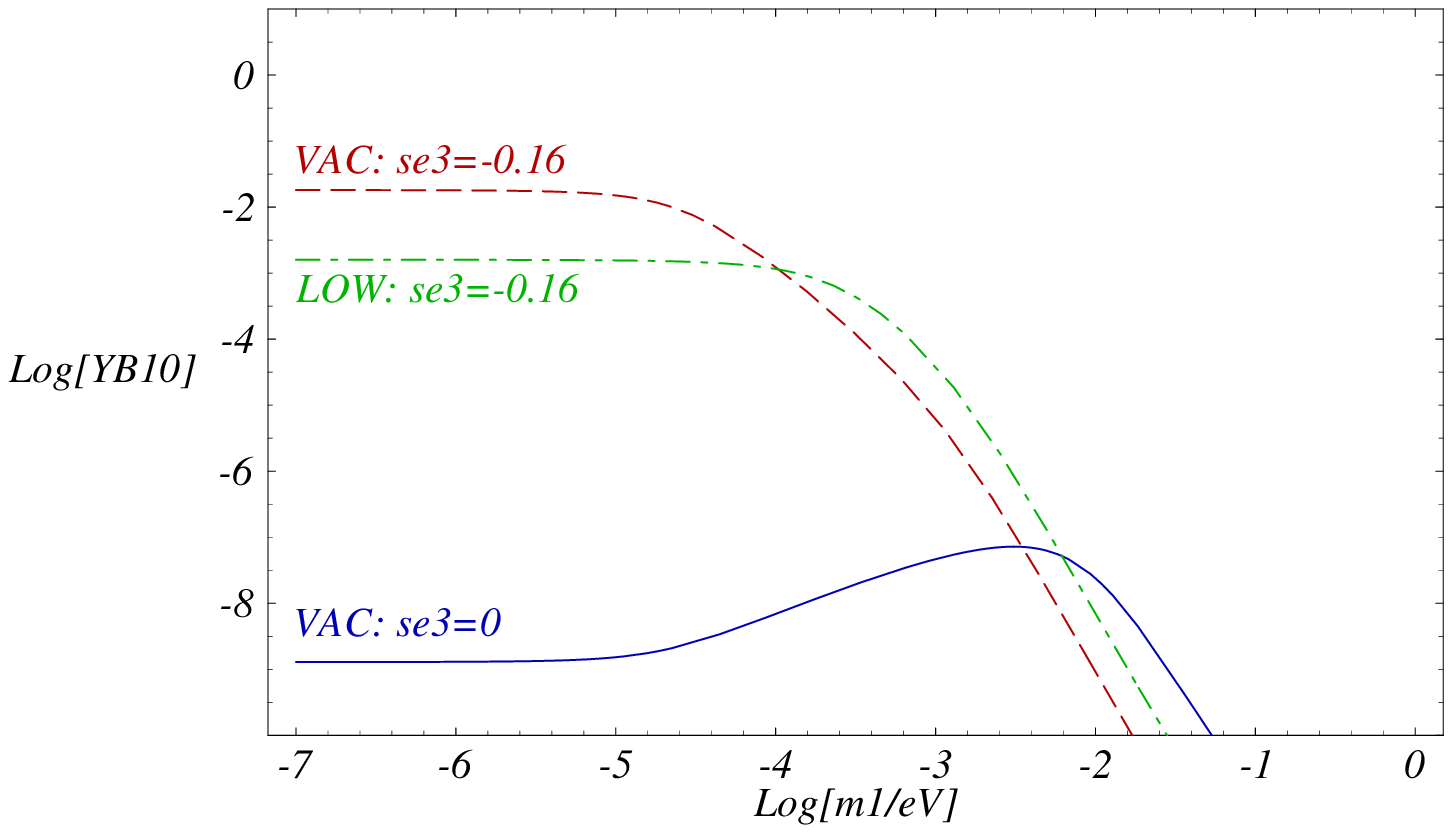}
\caption{The baryon asymmetry $Y_{B10}$ for LOW and VAC oscillations of solar 
  neutrinos. The plain curve is the vacuum case for vanishing $s_{e3}$.
  The gain from tuning the value $s_{e3}=-0.16$ is shown for the VAC
  (dashed) and LOW (dash-dotted) cases, taking for each case the
  maximising value of the phase $\delta_{13}$.}
\label{AsymVOLOW}
}

If we now turn on the largest allowed atmospheric and solar mixing, we get
the plain curves in figure~\ref{3MRVO}. Decreasing \( m_{1} \) from the
unaffected degenerate limit, \( M_{3} \) starts again levelling off at \(
m_{1}=\Delta m_{atm} \).  But the maximal atmospheric mixing immediately
induces the type of ``level crossing'' seen in the previous section, which
effectively exchanges \( M_{2} \) and \( M_{3} \). \( M_{2} \) thus levels
off at a much lower value \( \approx 10^{10}\mathrm{GeV} \) which offers a
better possibility for \( M_{1} \) to catch up. However, despite keeping \(
s_{e3}=0 \), CKM mixing in eq. (\ref{Ueff}) still induces a non-trivial \(
|s_{13eff}|\approx 0.16 \) which stops the growth of \( M_{1} \) below \(
m_{1}=|s_{13eff}|^{2}\Delta m_{atm}\approx 10^{-3}\mathrm{eV} \).  \( M_{2}
\) on the contrary starts growing again until it is hit by the solar mixing
at \( m_{1}\approx \Delta m_{sun} \). We thus expect a maximum baryon
asymmetry around \( m_{1}\approx 10^{-3}\mathrm{eV} \) in this case. A plot
of the full 3 flavour asymmetry (plain curve in figure~\ref{AsymVOLOW})
confirms this expectation, and further shows that this maximum is at least
5 orders of magnitude too low.  The shape of the asymmetry
nicely fits the picture derived in the previous section, once we recall that
for \( m_{1}<\Delta m_{sun} \) both \( M_{1} \) and \( M_{2} \) stay
constant. It is also worth noticing that at this stage, the only source of
CP violation is $\delta^{CKM}$, and that it drives the right sign for the
baryon asymmetry, albeit with a small amplitude.

\FIGURE{
\begin{tabular}{cc}
\psfrag{-1}[c][c]{\scriptsize -1}
\psfrag{-0.5}[c][c]{\scriptsize -0.5}
\psfrag{0}[c][c]{\scriptsize 0}
\psfrag{0.5}[c][c]{\scriptsize 0.5}
\psfrag{1}[c][c]{\scriptsize 1}
\psfrag{se3}[tc][Bc]{$s_{e3}$}
\psfrag{-3}[r][r]{\scriptsize -3}
\psfrag{-2}[r][r]{\scriptsize -2}
\psfrag{-1}[r][r]{\scriptsize -1}
\psfrag{0}[r][r]{\scriptsize 0}
\psfrag{1}[r][r]{\scriptsize 1}
\psfrag{2}[r][r]{\scriptsize 2}
\psfrag{3}[r][r]{\scriptsize 3}
\psfrag{d13}[r][r]{$\delta_{13}$}
\psfrag{Vacuum ; m1=1E-5 eV}[cb][cb]{Vacuum ; $m_1=10^{-5}$ eV}
\includegraphics[width=0.45\textwidth]{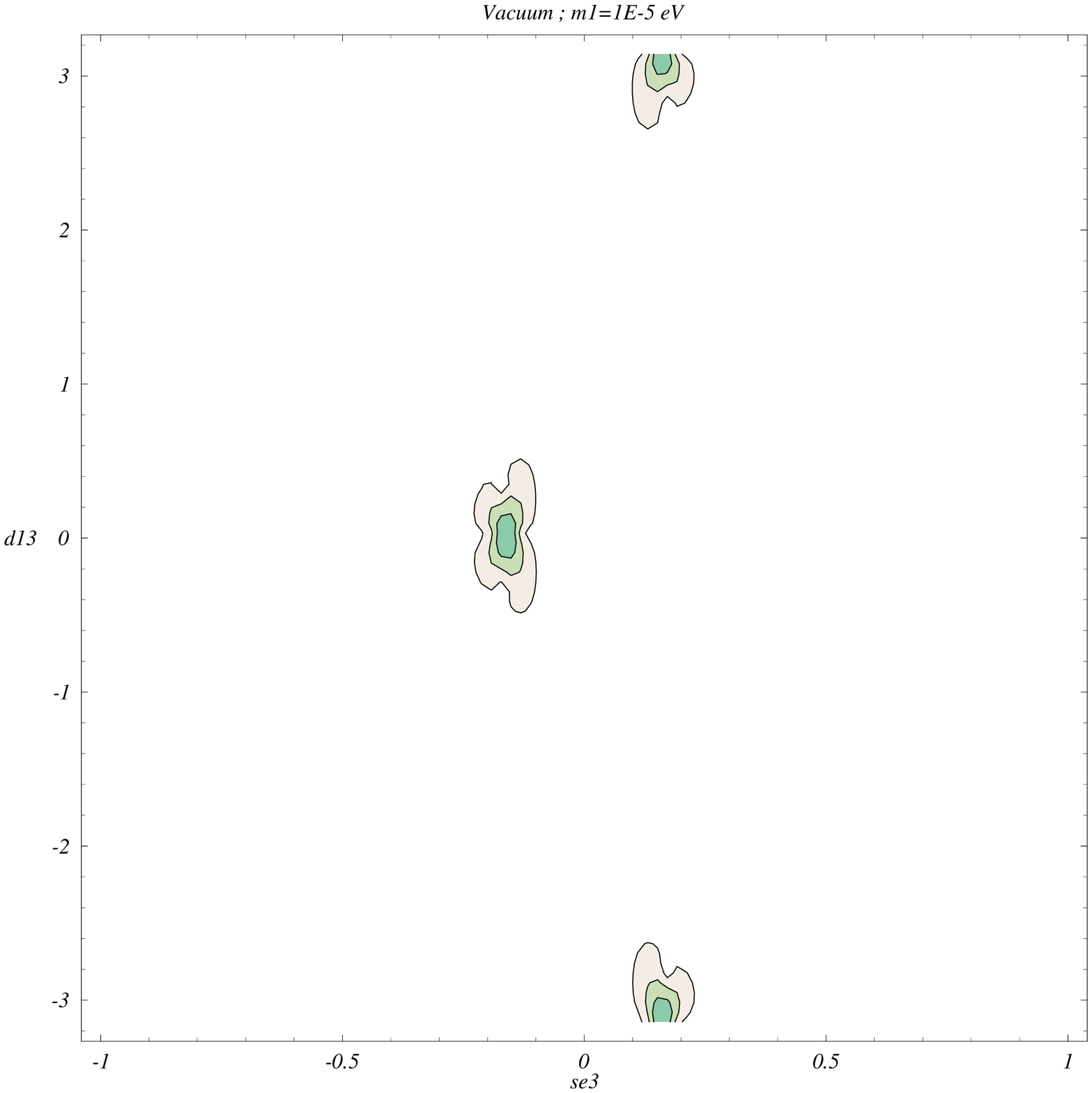}&
\psfrag{-0.18}[tc][tc]{{\scriptsize -0.18}}
\psfrag{-0.17}[tc][tc]{{\scriptsize -0.17}}
\psfrag{-0.16}[tc][tc]{\scriptsize -0.16}
\psfrag{-0.15}[tc][tc]{\scriptsize -0.15}
\psfrag{-0.14}[tc][tc]{\scriptsize -0.14}
\psfrag{se3}[tc][bc]{\raisebox{0pt}[2ex]{$s_{e3}$}}
\psfrag{-0.4}[r][r]{\scriptsize -0.4}
\psfrag{-0.2}[r][r]{\scriptsize -0.2}
\psfrag{0}[r][r]{\scriptsize 0}
\psfrag{0.2}[r][r]{\scriptsize 0.2}
\psfrag{0.4}[r][r]{\scriptsize 0.4}
\psfrag{d13}[r][r]{$\delta_{13}$}
\psfrag{Vacuum ; m1=1E-5 eV}[cb][cb]{Vacuum ; $m_1=10^{-5}$ eV}
\includegraphics[width=0.45\textwidth]{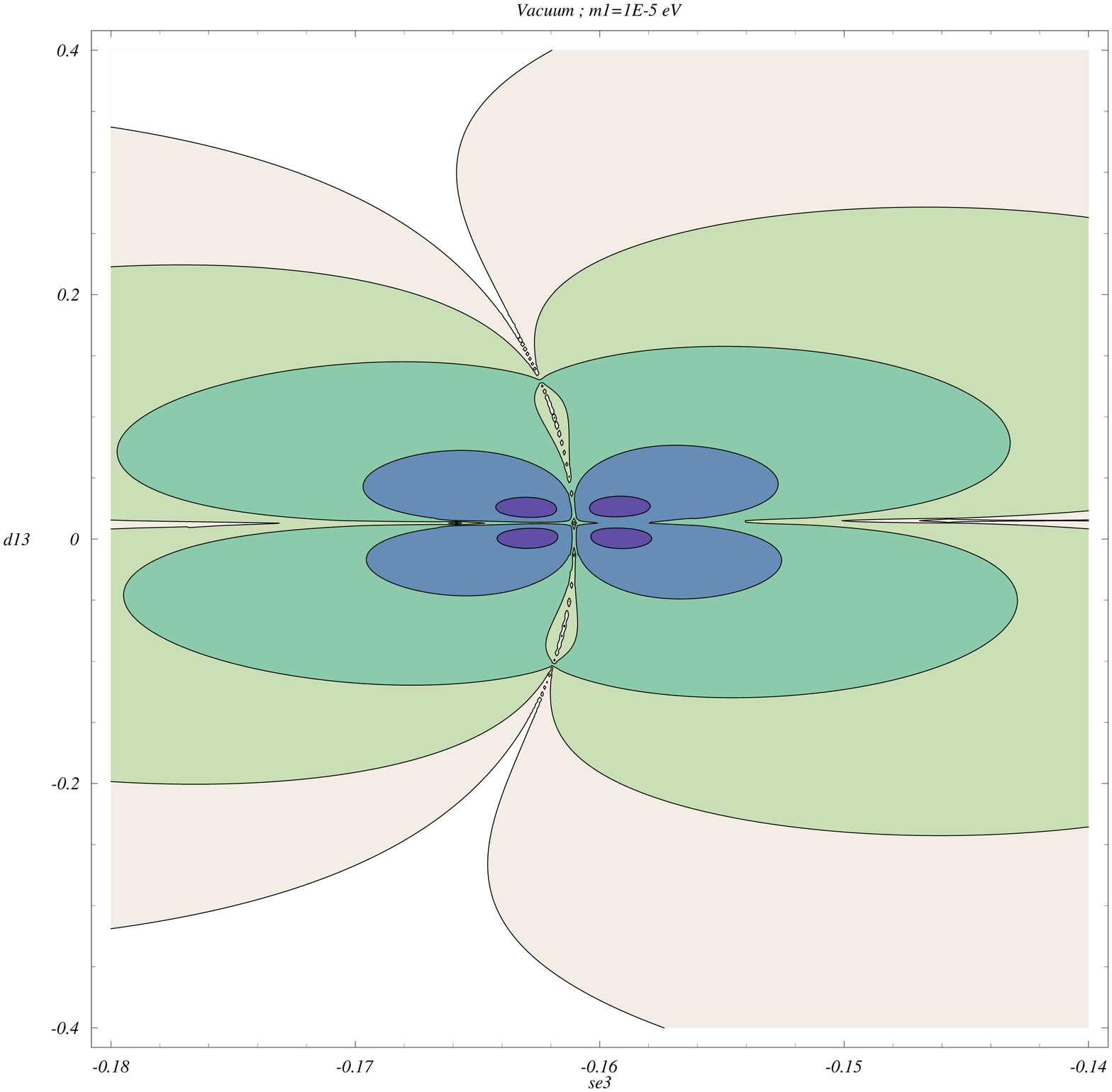}
\end{tabular}
\caption{Sensitivity of the baryon asymmetry on the matrix element
  $U_{e3}=s_{e3}e^{i\delta_{13}}$ most constrained by CHOOZ. The levels of
  $Y_{B10}$ shown are $10^{-6},\ 10^{-5}\ldots\ 10^{-2}$, the latter being only
  obtained on the darkest dots around $(s_{e3}= -0.16,\delta_{13}= 0.0)$
  at the centre of the zoom on the right.}
\label{butterfly}
}

To get larger results, we may use the freedom left by the CHOOZ experiment
to play with \( s_{e3} \). Indeed, for \( s_{e3}\approx -0.16 \) (dashed
curve in figure~\ref{3MRVO}), \( s_{13eff}\approx 0 \) so that \( M_{1} \)
is decoupled from the value of \( m_{3} \), and continues rising up to \(
m_{u}^{2}/9\Delta m_{sun}\approx 10^{8}\mathrm{GeV} \).  This may give a
rise in the asymmetry down to \( m_{1}\approx 10^{-5}\mathrm{eV} \).
Notice that for such a value, $K\approx 1$ which minimises dilution
effects, \( M_{1}\approx 10^{8}\mathrm{GeV} \) which is safe {\it w.r.t.}
gravitino bounds on inflation in SUSY versions of this scenario, and \(
M_{3}\approx 10^{16}\mathrm{GeV} \) which is marginally compatible with the
GUT scale.  It is worth detailing where the special value \( s_{e3}\approx
-0.22\, s_{atm} \) comes from. In the expression of \( U_{eff} \), we may
in a first approximation neglect all CKM mixing except the Cabbibo \(
V_{12}^{CKM} \).  Then keeping terms at most linear in \( s_{12}^{CKM} \)
and \( s_{e3} \), we may write:
\begin{eqnarray}
U_{eff} & \approx  & V_{12}^{CKM}\, U_{23}U_{13}U_{12}\\
 & = & U_{23}U_{13}\, 
      [U_{13}^{\dagger} U_{23}^{\dagger }\,V_{12}^{CKM}\,U_{23}U_{13}]\,
       U_{12}\\
 & \approx  & U_{23}U_{13}\,V'_{13}V'_{12}\,U_{12}\label{Ueffapprox} \\
 & \doteq  & U_{23eff}\, U_{13eff}\, U_{12eff}
\end{eqnarray}
with the commutator $ V'$ matrices parameterised like (\ref{VCKM}) with
\( s'_{12}\approx c_{atm}\, s_{12}^{CKM} \) and 
\( s'_{13}\approx s_{atm}e^{i\delta_{23}}\, s_{12}^{CKM} \). Cancelling 
\( s_{13eff} \) then
requires to satisfy the complex equation
\begin{equation} 
U_{e3}=s_{e3}e^{-i\delta_{13}}\approx 
    -s_{atm}e^{-i\delta_{23}}\, s_{12}^{CKM}= -0.16\,e^{-i\delta_{23}}
    \label{TuneUe3}
\end{equation}
which is illustrated for $\delta_{23}=0$ in figure~\ref{butterfly}. One
sees the extreme tuning on both the modulus and the phase of $U_{e3}$,
needed to achieve an enhancement in the four dark central dots of the
zoom. Even the sign of the asymmetry
violently flips in the four quadrants of this zoom.

\FIGURE{
\psfrag{-7}[c][c]{-7}
\psfrag{-6}[c][c]{-6}
\psfrag{-5}[c][c]{-5}
\psfrag{-4}[c][c]{-4}
\psfrag{-3}[c][c]{-3}
\psfrag{-2}[c][c]{-2}
\psfrag{-1}[c][c]{-1}
\psfrag{0}[c][c]{0}
\psfrag{Log[m1/eV]}[tc][c]{$Log_{10}[m_1/{\rm eV}]$}
\psfrag{-8}[r][r]{-8}
\psfrag{-6}[r][r]{-6}
\psfrag{-4}[r][r]{-4}
\psfrag{-2}[r][r]{-2}
\psfrag{0}[r][r]{0}
\psfrag{2}[r][r]{2}
\psfrag{Log[YB10]}[bc][r][1][90] {$Log_{10}[Y_{B10}]$}
\psfrag{se3=-0.16, ssun=-0.157, d13=0.005}[bl][bl]{{\darkgreen \scriptsize 
    $s_{e3}=-0.16$, $s_{sun}=-0.157$, $\delta_{13}=0.005$}}
\psfrag{SMA: se3=-0.156, d13=-0.02}[tl][tl]{{\blue \scriptsize 
    $\begin{array}{ll} \rm{SMA: }&s_{e3}=-0.156,\\
      &\delta_{13}=-0.02
    \end{array}$}}
\psfrag{LMA: se3=-0.12, d13=-0.48}[bl][l]{{\darkred \scriptsize 
    LMA: $s_{e3}=-0.12$, $\delta_{13}=-0.48$}}
\psfrag{LMA: se3=0}[bl][l]{{\darkred \scriptsize LMA: $s_{e3}=0$}}
\psfrag{SMA: se3=0}[tl][l]{{\blue \scriptsize SMA: $s_{e3}=0$}}
\includegraphics[width=0.75\textwidth]{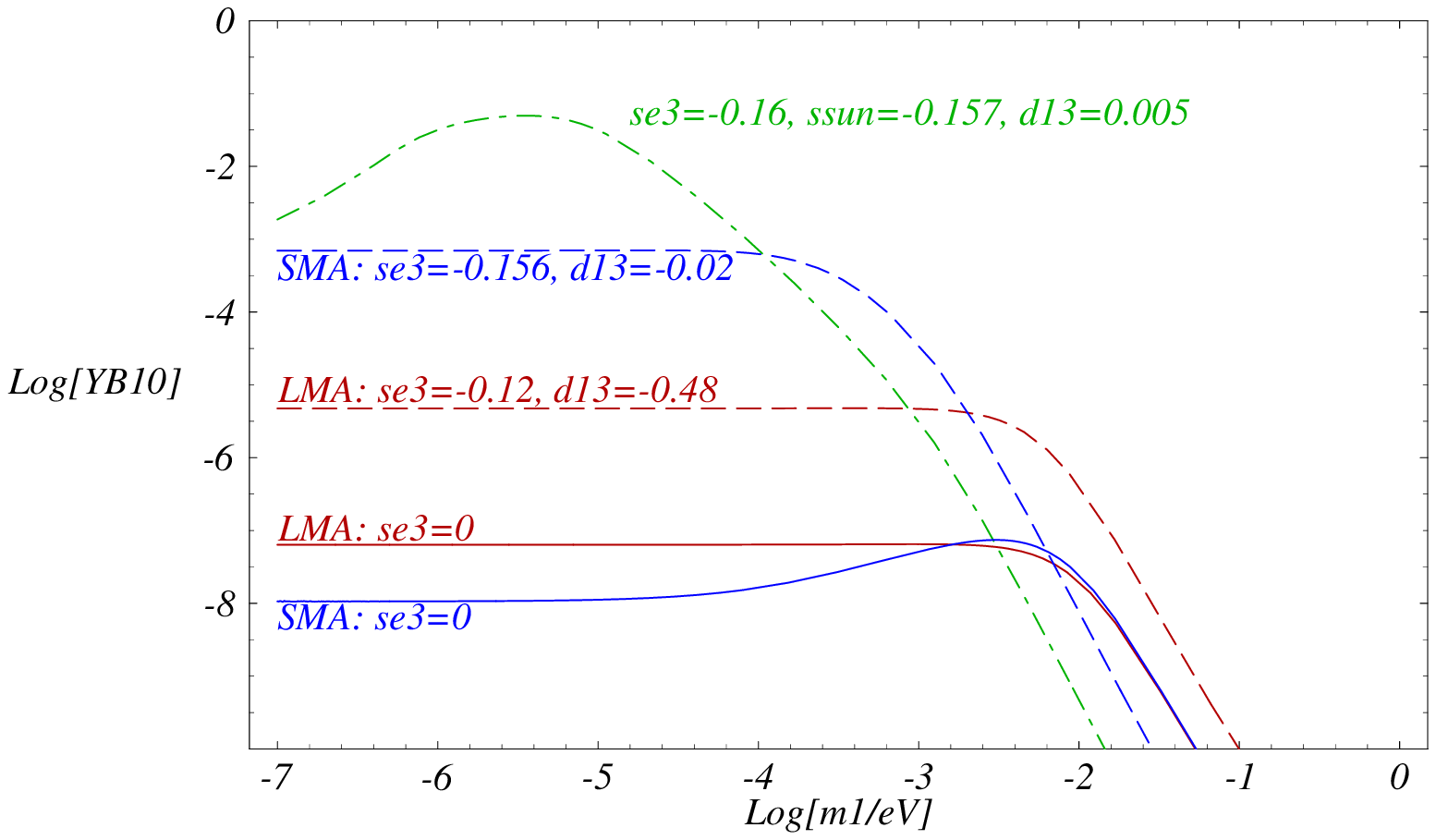}
\caption{The baryon asymmetry $Y_{B10}$ for MSW oscillations of solar 
  neutrinos with various mixing angles. The plain curves are for vanishing
  $s_{e3}$ with no phase except $\delta^{CKM}$. The dashed SMA (resp. LMA)
  curves are for $\delta_{13}=-0.02$ (resp. $-0.1$) .  Only the particular
  dash-dotted curve ($s_{e3}=-0.16,\ s_{sun}=-0.156,\ \delta_{13}=0.005$)
  can nearly produce enough asymmetry, but it can not account for the solar
  neutrino deficit.}
\label{AsymMSW}
}

The previous discussion can be repeated for the LOW solution, but the
larger value of $\Delta m_{sun}^2$ leaves less room for the asymmetry to
grow before levelling off. As seen in figure \ref{AsymVOLOW}, the maximum
that can hoped for is 2 orders of magnitude too low.

If we now turn to MSW solutions of the solar neutrino deficit, we get a
still larger value of \( \Delta m_{sun} \), and there is too little room
for \( M_{1} \) to grow between \( m_{1}\approx \Delta m_{atm} \) and \(
m_{1}\approx \Delta m_{sun} \).  However, it can grow for smaller \( m_{1}
\), provided both \( s_{13eff}=s_{12eff}=0 \). This is in principle
possible, but now requires very special values for both \( s_{e3}\approx
-0.16 \) (as previously) and \( s_{sun}\approx -s_{12}^{CKM}\,
c_{atm}\approx -0.16 \) (see eq.  \ref{Ueffapprox}).  This last value of \(
s_{sun} \) is incompatible with any possible MSW solution, unless taking
$s_{12}^{CKM}$ away from its quark-lepton symmetry value, which is
illustrated as a function of \( m_{1} \) and \( s_{sun} \) on figures
\ref{AsymMSW} and \ref{AsymMSWssun}.

\FIGURE{
\psfrag{-1}[c][c]{-1}
\psfrag{-0.5}[c][c]{-0.5}
\psfrag{0}[c][c]{0}
\psfrag{0.5}[c][c]{0.5}
\psfrag{1}[c][c]{1}
\psfrag{ssun}[c][cb]{$s_{sun}$}
\psfrag{-10}[r][r]{-10}
\psfrag{-8}[r][r]{-8}
\psfrag{-6}[r][r]{-6}
\psfrag{-4}[r][r]{-4}
\psfrag{-2}[r][r]{-2}
\psfrag{0}[r][r]{0}
\psfrag{Log[YB10]}[c][c][1][90]{$Log_{10}[Y_{B10}]$}
\psfrag{SMA}{{\blue SMA}}
\psfrag{LMA}{{\darkred LMA}}
\includegraphics{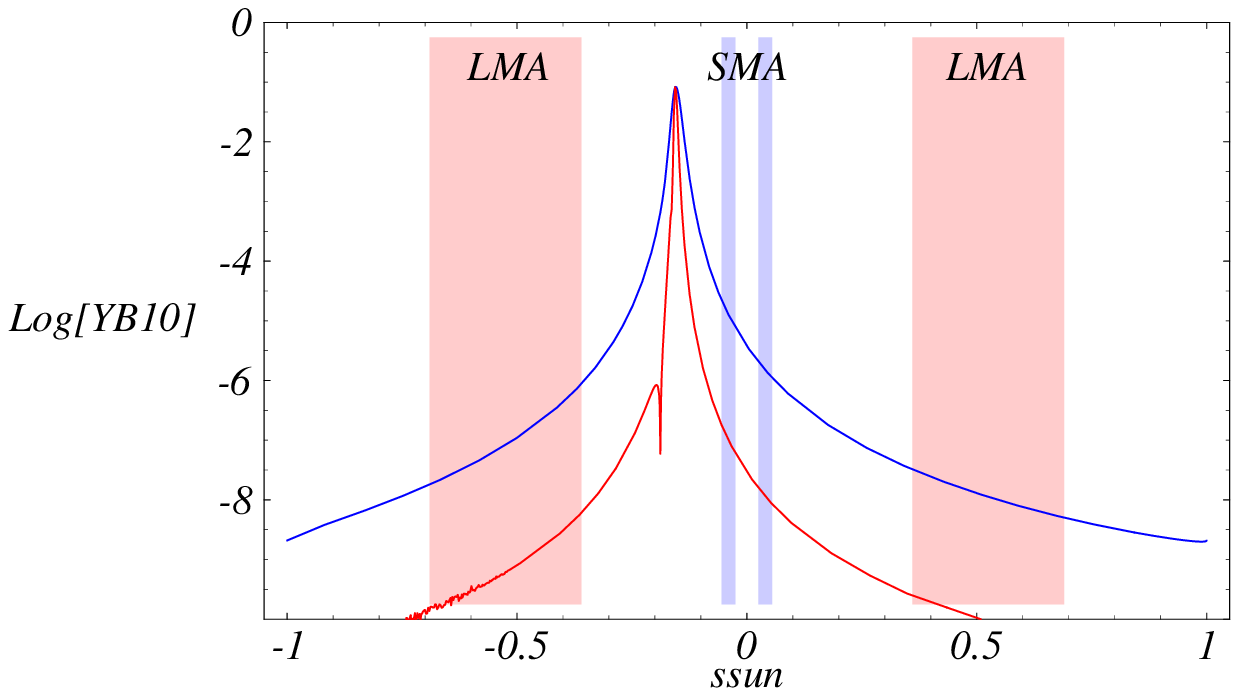}
\caption{The baryon asymmetry $Y_{B10}$ for MSW oscillations of solar
  neutrinos (LMA: lower red, SMA: upper blue), as a function of $s_{sun}$.
  The maximum is nearly large enough (and sharply peaked) for $m_{1}\approx
  10^{-5}$ if $s_{e3}=-0.16$, but lies just between the small and large
  mixing angle rectangles.}
\label{AsymMSWssun}
}

With a standard mass hierarchy, we thus see that vacuum oscillations
are the closest to account both for the solar neutrinos deficit and the baryon
asymmetry of the universe. With the same line of reasoning, it is easy to
see that with inverted hierarchy (\( m_{3}\ll m_{1}\ll m_{2}) \) the
asymmetry is even smaller, because of a larger \( M_{1}-M_{2} \) gap coming
from the bound \( M_{1}< m_{u}^{2}/6\Delta m_{atm} \).  Indeed, as the
free parameter \( m_{3} \) is lowered below \( \Delta m_{atm} \), two \(
M_{R} \) eigenvalues must now stop growing instead of one. There are then
two possibilities: if \( M_{1} \) is the only one that grows in the absence
of mixing, then maximal atmospheric mixing stops its growth once \(
m_{3}<s_{atm}\Delta m_{atm} \).  If on the other hand \( M_{1} \) stops
growing without mixing, then mixing won't help, as it can only induce
``levels repulsion'', not attraction. In both cases, \( M_{1} \) is at most
\( 10^{4}\mathrm{GeV} \), and \( M_{1}/M_{2} \) cannot exceed \(
m_{u}^{2}/m_{c}^{2}\approx 10^{-6} \) which makes the asymmetry too small.

Let us now review the effect of the various phases for trying to reach a
sufficient asymmetry. In the discussion above, we have only switched on the
minimal phases necessary to get a non-trivial result, namely $\delta^{CKM}$
which is unavoidable in our quark-lepton symmetry set-up, and $\delta_{13}$
which was necessary to enforce the cancellation (\ref{TuneUe3}). From this
equation, it is clear that if $\delta_{23}$ were non zero, it would merely
shift the optimal value of $\delta_{13}$, the asymmetry being a function of
the combination $\delta_{13}-\delta_{23}$ in the region of the low $m_1$
peak. An non-zero $\delta_{12}$ would induce a much smaller shift, through
the sub-leading corrections to (\ref{TuneUe3}). Less trivial are the
effects of the Majorana phases $\phi_{1,2}$. As suggested by equation
(\ref{YBcorrect}), these can lead to cancellations and near degenerate
right-handed masses for a fine tuned value of one of the phases and of
$m_1$. However, the peak in the decay asymmetries $\epsilon_{1,2}$ and
$\epsilon_{1,3}$ tend to cancel each other, and get further reduced by a
similar peak in $K_1$, so that no enhancement of the final asymmetry can be
produced for the large $m_1\approx 10^{-3}$eV where it could occur.

\section{Conclusions and perspectives \label{sec:concl}}

In this work, we have studied the possible relations between neutrino
oscillations and the baryon asymmetry of the universe produced by
equilibrium decays of right handed neutrinos, assuming an \( SO(10) \)
see-saw mechanism for neutrino masses.  As explained, this is a minimal
predictive set of assumptions to study such relations. We find that the
produced asymmetry is generically six orders of magnitude too small because
of the huge Dirac masses hierarchy assumed (\(
r^{2}=m_{u}^{2}/m_{c}^{2}\approx 10^{-6} \)).  The closest one can come to
the observed asymmetry in this framework is at least one order of magnitude
too small and requires vacuum oscillations of solar neutrinos, with
standard hierarchy and very specific values for the least constrained
neutrino parameters: 1) the lightest mass (\( m_{1}\approx \Delta
m_{sun}\approx 10^{-5}\mathrm{eV} \)), and 2) the heavy component of the
electron neutrino (\( |s_{e3}|=|s_{atm}|.|\sin (\theta _{Cabbibo})|\approx
0.16 \)).

This result calls for several extensions and refinements. It would first of
all be more satisfactory to get the right-handed mass relations we tend to
need out of some theoretical mass model for all particles. This seems quite
challenging at the moment.  Nice family symmetry models like
\cite{Irges:1998ax,Sato:1998hv} for instance, generically have a much
larger \( M_{1}/M_{2} \) hierarchy than we found necessary for leptogenesis.

Second, one may question our crude implementation of the \( SO(10) \)
relations between quarks and leptons Dirac-masses. We simply used \(
m_{lepton}=m_{quark}/3 \), which fails by about a factor 3 for the
muon-strange mass ratio. Since the asymmetry goes like \( r^{2} \) (see eq.
\ref{YBapprox}), such small factors should not qualitatively change our
conclusions; taking 10 instead of 1MeV for the lowest neutrino Dirac mass
could for instance barely bring the vacuum oscillation solution up to the
required asymmetry $Y_{B10}\approx1$.  The correct running of neutrino
Yukawa couplings may have important effects on the mixing, which could
change the interesting value of \( s_{e3} \), and for instance make it
inaccessible to first generation long baseline experiments, or on the
contrary, already excluded by CHOOZ.

Our conclusions seemingly contradict previous literature on the subject,
which usually left the impression that leptogenesis could work without
restrictions.  This comes from our maybe over-restrictive framework,
insisting on both \( SO(10) \) relations, and recent neutrino oscillations
data. Let us review some of these discrepancies. In reference
\cite{Buchmuller:1996pa,Plumacher:1997ru}, the authors use \( SO(10)
\)-inspired see-saw relations like eq. \ref{MRseesaw} and conclude that
leptogenesis is generically possible.  However, the values taken in the
final analysis are hard to reconcile with maximal atmospheric mixing, which
at the time was not as firmly established as today. In subsequent works
\cite{Buchmuller:1998zf,Buchmuller:1999cu}, maximal atmospheric mixing is
well taken into account, but the Dirac mass pattern considered is derived
from a family symmetry in an \(SU(5)\) framework; this gives typically larger
``CKM'' mixing and much weaker Dirac hierarchy than the simple quark-lepton
ansatz we considered, both of which increase the asymmetry.  In reference
\cite{Goldberg:1999hp}, it is shown that leptogenesis can be reconciled
with SMA solar neutrinos. But the Dirac masses needed badly violate \(
SO(10) \) relations. Finally, another way to loosen the strong \( SO(10) \)
constraints is to invoke what could be coined ``see-saw in the scalar
sector'', namely a small v.e.v. for a scalar triplet that directly
contributes to the left neutrino Majorana mass \cite{Mohapatra}. This
however introduces a new parameter, and non-trivial constraints on the
scalar sector as in \cite{Joshipura:1999is}.

\paragraph{Note added:} While this work was being revised, others 
\cite{Falcone:2000ib},\cite{Branco:2002kt} have confirmed the important
effect of $U_{e3}$ in the framework discussed here.  It should however be
stressed that the type of random scan used in these works requires a very
fine graining to hit the maximal possible value: at least $10^5$ points in
each $U_{e3}$ plane, as seen figure~\ref{butterfly}. After the recent
KamLAND results\cite{Eguchi:2002dm}, the only surviving LMA solution requires an
optimal value of $|U_{e3}|\sim 0.12$ which is but slightly lower than the VAC
case detailed in this work. However, reconciling thermal leptogenesis with an
$SO(10)$ mass pattern seems even harder than before.

\subsubsection*{\bibliographystyle{JHEP}
  \bibliography{leptosc-6} }

\end{document}